# Front-end electronic readout system for the Belle II imaging Time-Of-Propagation detector

Dmitri Kotchetkov[a], Oskar Hartbrich[a,*], Matthew Andrew[a], Matthew Barrett[a,1], Martin Bessner[a], Vishal Bhardwaj[b,2], Thomas Browder[a], Julien Cercillieux[a], Ryan Conrad[c], Istvan Danko[d], Shawn Dubey[a], James Fast[c], Bryan Fulsom[c], Christopher Ketter[a], Brian Kirby[a,3], Alyssa Loos[b], Luca Macchiarulo[a], Boštjan Maček[a,4], Kurtis Nishimura[a], Milind Purohit[b], Carl Rosenfeld[b], Ziru Sang[a], Vladimir Savinov[d], Gary Varner[a], Gerard Visser[e], Tobias Weber[a,5], Lynn Wood[c]

a. Department of Physics and Astronomy, University of Hawaii at Manoa, 2505 Correa Road, Honolulu, HI 96822, USA
b. Department of Physics and Astronomy, University of South Carolina, 712 Main Street, Columbia, SC 29208, USA
c. Pacific Northwest National Laboratory, 902 Battelle Boulevard, Richland, WA 99354, USA
d. Department of Physics and Astronomy, University of Pittsburgh, 3941 O'Hara Street, Pittsburgh, PA 15260, USA
e. Center for Exploration of Energy and Matter, Indiana University, 2401 North Milo B. Sampson Lane, Bloomington, IN 47408, USA

* Corresponding author. Phone: 1-808-956-4097. Fax: 1-808-956-2930
E-mail address: ohartbri@hawaii.edu

[1]Present address: Department of Physics and Astronomy, Wayne State University, 666 W Hancock St, Detroit, MI 48201, USA
[2]Present address: Indian Institute of Science Education and Research Mohali, Knowledge city, Sector 81, SAS Nagar, Manauli PO 140306, India
[3]Present address: Physics Department, Brookhaven National Laboratory, Upton, NY 11973, USA
[4]Present address: Department of Experimental Particle Physics, Jožef Stefan Institute, Jamova cesta 39, 1000 Ljubljana, Slovenia
[5]Present address: Ruhr-Universität Bochum, 44780 Bochum, Germany

**Abstract**

The Time-Of-Propagation detector is a Cherenkov particle identification detector based on quartz radiator bars for the Belle II experiment at the SuperKEKB $e^+e^-$ collider. The purpose of the detector is to identify the type of charged hadrons produced in $e^+e^-$ collisions, and requires a single photon timing resolution below 100 picoseconds. A novel front-end electronic system was designed, built, and integrated to acquire data from the 8192 microchannel plate photomultiplier tube channels in the detector. Waveform sampling of these analog signals is done by switched-capacitor array application-specific integrated circuits. The processes of triggering, digitization of windows of interest, readout, and data transfer to the Belle II data acquisition system are managed by Xilinx Zynq-7000 programmable system on a chip devices.

1. Introduction

The Belle II experiment [1] at the SuperKEKB electron-positron collider (High Energy Accelerator Research Organization, KEK, Tsukuba, Japan) is an upgrade of the Belle experiment [2-3] that studied CP-violation, weak interaction coupling constants and rare physics processes at the Y(4S) and Y(5S) resonances, and completed data taking in 2010. SuperKEKB collides 7 GeV electron beams with 4 GeV positron beams, with an instantaneous design luminosity of $8 \times 10^{35}$ $cm^{-1}s^{-1}$ and a goal of recording 50 $ab^{-1}$ of integrated luminosity. Such large data samples will allow measurements of rare B and D meson decays, including those that are suppressed or forbidden by the Standard Model of particle physics. Belle II will also allow unprecedented sensitivity to lepton flavor violating decays of the $\tau$ lepton. In addition to searches for new physics, such data samples will lead to a substantial reduction of uncertainties for processes that were already measured by the previous generation of B-factory experiments. To detect rare processes, as well as to maximize the signal to background ratios in the channels of interest, Belle II requires improved particle identification capabilities. In particular, it is expected that rare and previously unobserved physics phenomena can be explored at Belle II if separation of kaons from pions in the transverse momentum range from 1 GeV/c to 4 GeV/c can be accomplished with 85-90% efficiency while the misidentification rate is maintained below 5% [1, 4]. Improved particle identification performance is also needed for Belle II to minimize the effects of beam backgrounds expected from SuperKEKB. To meet such particle identification requirements in the barrel region of Belle II, a novel Cherenkov radiation detector – the Time-Of-Propagation (TOP) detector [5-8] – was built (Fig. 1).

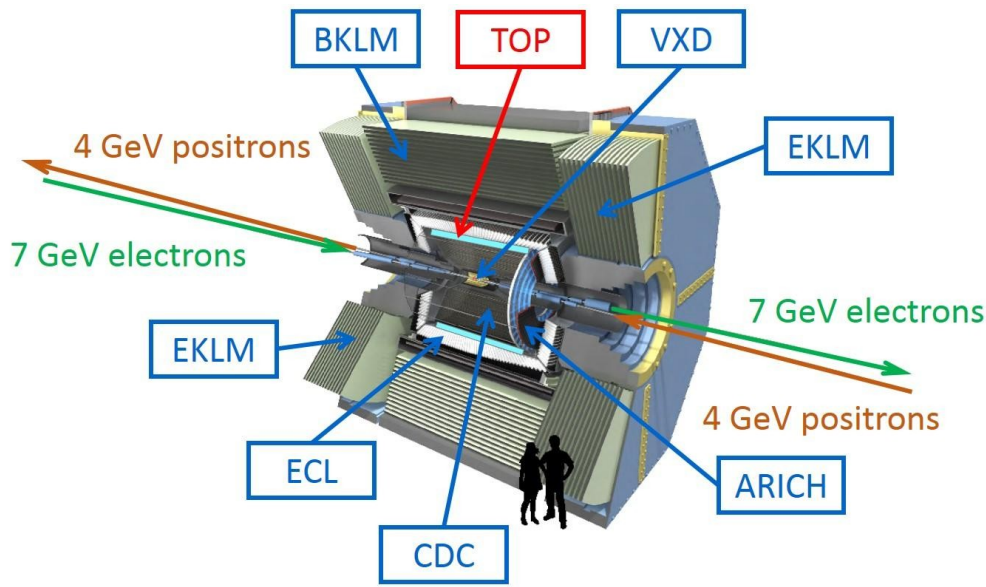

*Figure 1*: *Location of the Time-Of-Propagation detector in Belle II. VXD: Vertex Detector. CDC: Central Drift Chamber. ECL: Electromagnetic Calorimeter. ARICH: Aerogel Ring-Imaging Cherenkov Detector. BKLM: Barrel Kaon-Long and Muon detector. EKLM: Endcap Kaon-Long and Muon detector.*

2. Particle identification with TOP

When a relativistic charged particle passes through a dielectric medium with velocity above the speed of light in the said medium, the incident particle will generate a trail of optical photons due to the Cherenkov effect. The emission angle of this cone of Cherenkov photons depends on the velocity of the charged particle and the refractive index of the radiator material. For a given momentum, each hadron species (e.g. pions, kaons, protons) emits photons with a characteristic Cherenkov angle, which can thus be used to reconstruct the species of the incident particle.

In TOP, quartz (fused silica) bars are used as the radiator material for their high refractive index and excellent optical transparency. Due to the high critical angle of the quartz-air interface, Cherenkov photons generated inside the quartz radiator can be captured inside and propagate through the bar by total internal reflection, preserving their angular information. Such photons are then detected by a sensor array on one end of the quartz bar. These photons have different path lengths inside the quartz bar, and thus different times of propagation until they arrive at the sensor plane, depending on their Cherenkov angle (Fig. 2).

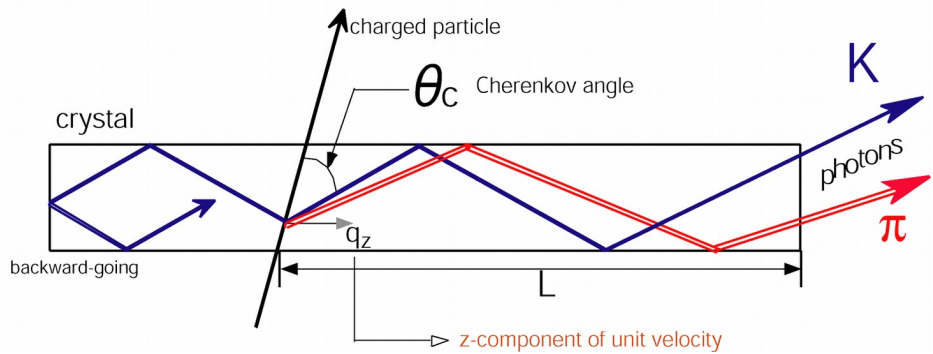

*Figure 2: Schematic illustration of the photon propagation in a TOP quartz bar. The path of propagation of Cherenkov photons from a kaon is shown in blue. The path of propagation of Cherenkov photons from a pion is shown in red. Due to the larger Cherenkov angle, photons originating from an incident pion have a shorter path length and thus (on average) earlier arrival times than photons from incident kaons.*

With knowledge of the charged particle momentum and of the location of the impact point on the quartz bar measured by the Central Drift Chamber, the TOP reconstruction algorithm can determine the species of an incident particle using the spatial and temporal distribution of the Cherenkov photons detected on the sensor plane. Since the propagation time is measured with respect to the well known reference time of the particle collision determined by the high precision radio-frequency clock of the SuperKEKB accelerator, the time-of-flight of the incident charged particle is implicitly included in the measurement and reconstruction. In order to reach the required particle identification performance goals, individual Cherenkov photons must be detected with a timing of better than 100 ps and a spatial resolution of a few millimeters on the sensor plane [4-5].

The TOP detector system is divided into 16 modules, cylindrically arranged between the Electromagnetic Calorimeter and the Central Drift Chamber of the Belle II detector. Each module consists of a radiator bar 20 mm thick and 450 mm wide (Fig. 3). The bar is glued from two identical 1250 x 450 x 20 mm$^3$ pieces and has a a total length of 2500 mm.

Photons that initially propagate away from the sensor plane are reflected back towards the sensor plane by a spherical mirror of 100mm length with a radius of curvature of 6500 mm, which is glued onto the end of the radiator. The spherical shape of the mirror effectively compensates the thickness of the bar by projecting reflected photons onto a single point on the image plane, depending on their angle of incidence.

A quartz prism is glued to the other end of the bar, which serves as an expansion volume for captured Cherenkov photons and widens their geometric distribution on the sensor plane. The prism has a length of 100 mm, a width of 456 mm, a thickness at the end of the bar of 20 mm, and a prism angle of 18.07 degrees.

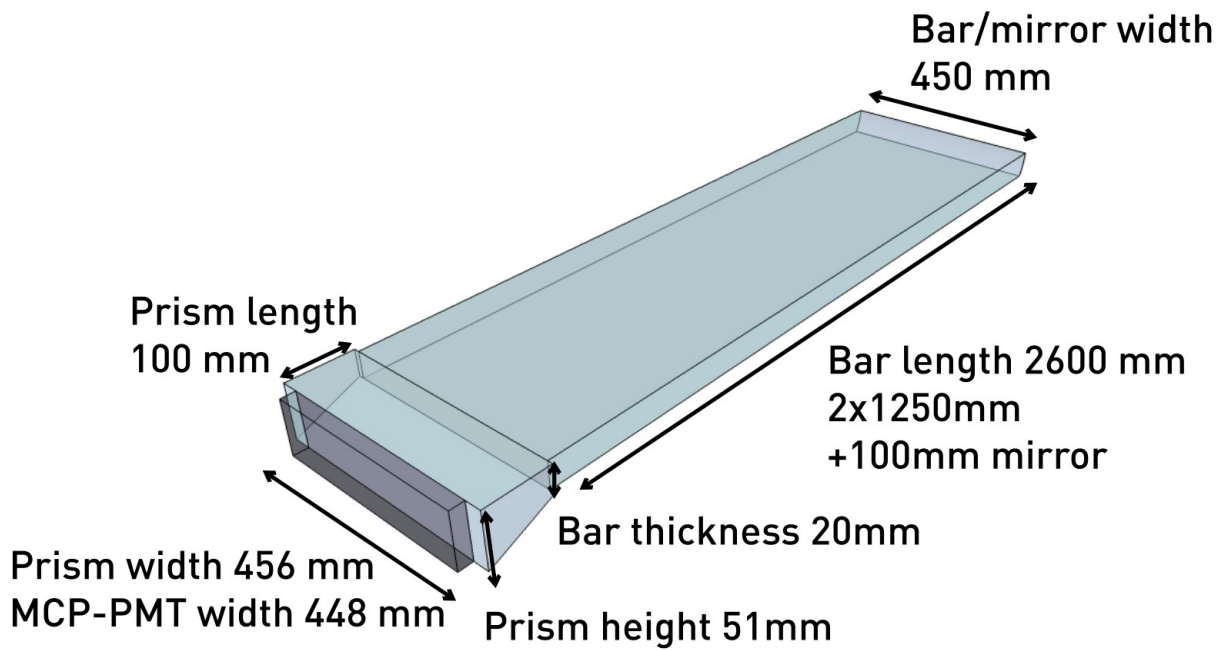

*Figure 3: Dimensions of the TOP module quartz components.*

For each quartz bar, two rows of 16 Hamamatsu R10754-07-M16(N) microchannel plate photomultiplier tubes (MCP-PMTs) [9, 10] (Fig. 4) are coupled to the outer surface of the prism to collect the Cherenkov photons. Every MCP-PMT has 16 readout anodes arranged in a 4 x 4 matrix of 5.275 x 5.275 mm$^2$ pixels, providing the required readout granularity.

A single normally incident charged hadron, either kaon or pion, with a momentum in the range 1 GeV/c to 2 GeV/c produces about 370 optical Cherenkov photons in one TOP bar, of which only 20-25 are detected due to the collection efficiency of the quartz, losses from the geometric acceptance of the photosensors, their quantum efficiencies, etc.

Each TOP module is integrated into a metal bar box housing the quartz bars, sensors and front-end electronics, which provides structural support, mechanical mounting points, geometric alignment and cooling services to the readout side (Fig. 5).

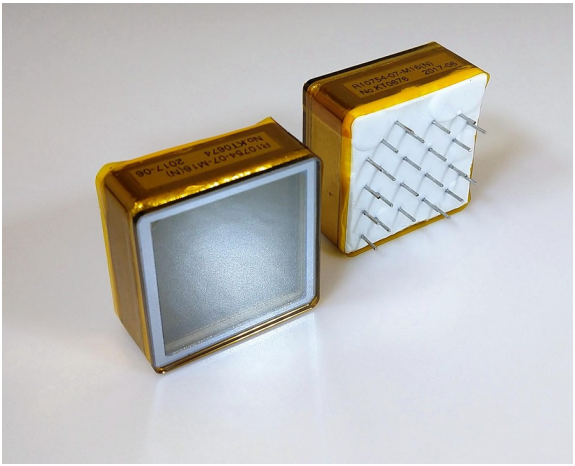

*Figure 5: 16-pixel Hamamatsu R10754-07-M16(N) microchannel plate photomultiplier tube. The dimensions are 27.6 x 27.6 x 13.1 mm3 without pin wires and 27.6 x 27.6 x 16.7 mm3 with pin wires.*

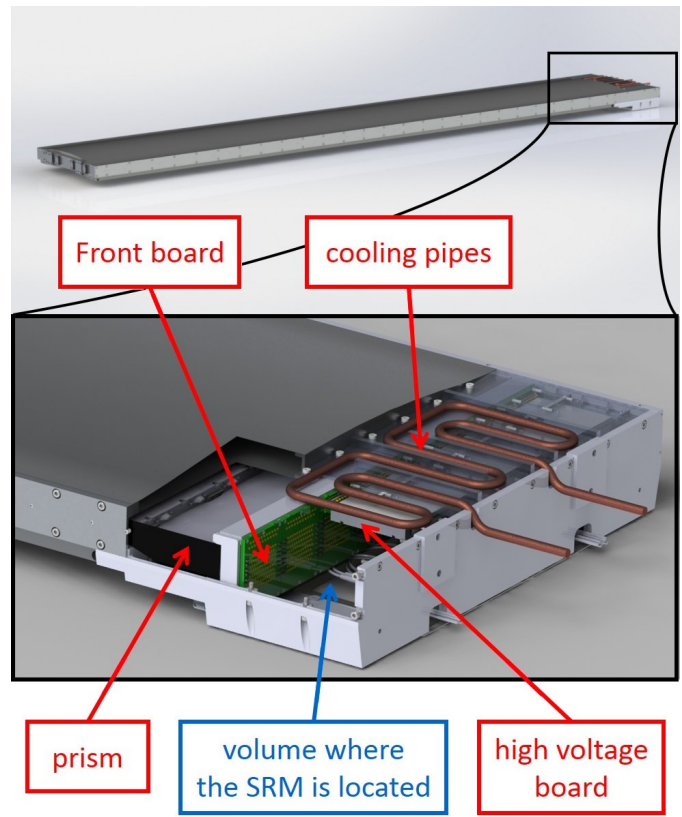

*Figure 4: Rendering of a TOP module bar box and detail cutaway view of the readout electronics installation space at the end of the bar box. The MCP-PMT sensors are mounted on the front board, facing towards the prism. The bar box enclosure is around 2850mm long and around 45mm wide*

3. Subdetector Readout Modules

The TOP front-end electronics are required to read out the signals of all 8192 MCP-PMT channels in the whole TOP system with a single photon timing resolution of better than 100 ps at a nominal trigger rate of up to 30kHz at the full projected luminosity of the SuperKEKB accelerator. This is achieved by employing specially designed fast waveform sampling electronics. Since the raw data rate of the waveform samples would overwhelm the transfer capabilities of the Belle II data acquisition (DAQ) system, all recorded waveforms need to be processed inside the detector front-end, so only the timing and pulse parameters of observed photons are transferred out of the detector.

The TOP readout system is organized as an ensemble of 64 compact standalone Subdetector Readout Modules (SRMs) as shown in Figure 6. Every quartz bar enclosure is equipped with four SRMs, each of which handles the readout of eight MCP-PMTs, corresponding to a total of 128 readout channels per SRM.

Mechanically, each SRM is a stack of five printed circuit boards. Four identical ASIC Carrier Boards amplify and read out the signals of 32 MCP-PMT anodes each. Every ASIC Carrier Board is equipped with four 8-channel custom-designed waveform sampling "Ice Ray Sampler version X" (IRSX) application-specific integrated circuits (ASICs). The name is derived from earlier waveform sampler designs for neutrino detection experiments in Antarctica [11, 12]. The waveform segments acquired on all four Carrier Boards are transferred to a single Standard Control Read-Out Data (SCROD) data aggregator board per SRM. The SCROD extracts the timing of photon pulses in the waveforms and transfers this feature extracted data to the off-detector electronics and the Belle II DAQ system. Figure 7 shows a fully assembled SCROD and Carrier Board.

A schematic diagram of the TOP data flow from the sensor through an SRM to the Belle II data acquisition system is shown in Figure 8.

The following sections discuss the individual components of the TOP readout electronics and their performance and calibration in greater detail.

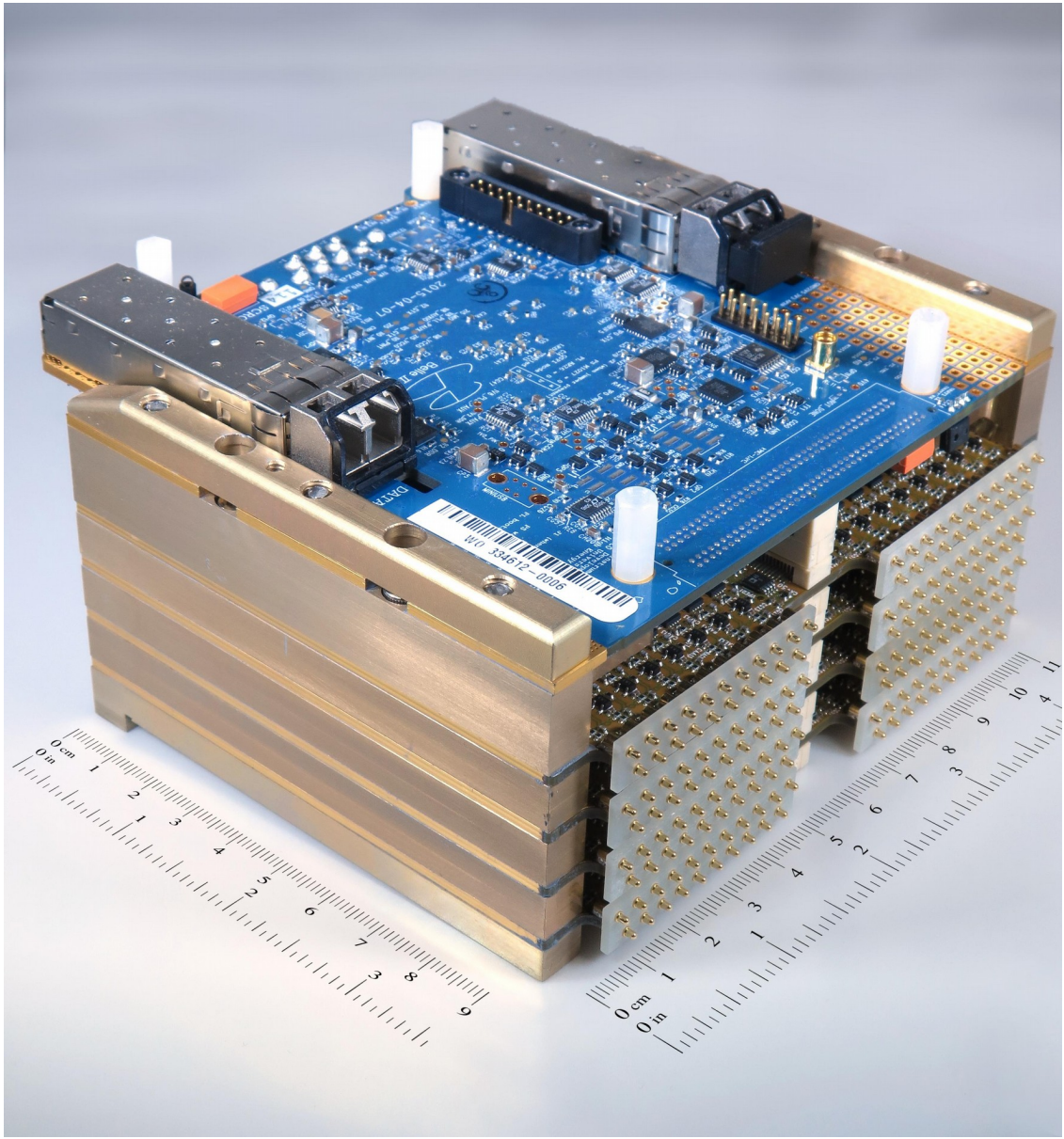

*Figure 6: TOP SRM consisting of one SCROD board (top) and four ASIC Carrier boards mounted as a board stack. The gold plated pogo-pins seen on the right connect to the MCP-PMT anodes through the Front Board. (Figure 15 shows a view with installed PMTs)*

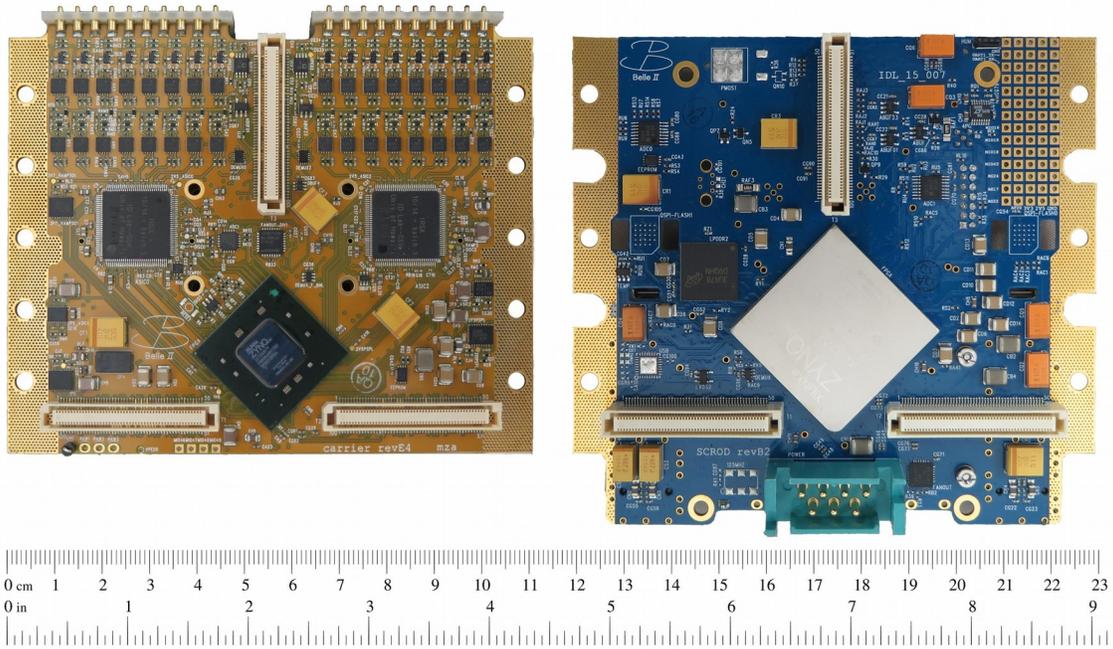

*Figure 8: Bottom view of TOP ASIC Carrier board (left) and SCROD board (right).*

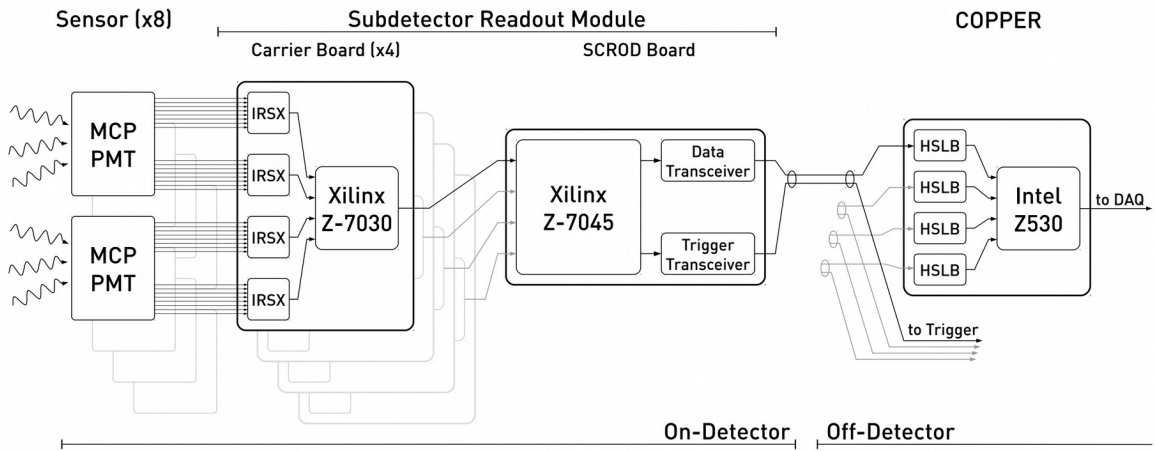

*Figure 7: Schematic view of the data flow through a TOP subdetector readout module. The MCP-PMT sensors on the left detect incident photons. The IRSX ASICs mounted on the four Carrier Boards sample the output of the sensor channels and convert the acquired analog waveforms to a digital representation. This digitised waveform is read out by the Z-7030 controller SoC mounted on each Carrier Board and transferred to the attached SCROD Board via an SRM-internal serial link. The SCROD Board is equipped with a Z-7045 SoC which processes the waveforms and sends out sparsified event data via ~20m long optical data links to the off-detector COPPER boards. Each COPPER board receives data from four TOP SRMs (corresponding to a whole quartz bar module) and forwards the data to the Belle II DAQ system.*

4. Analog signal sampling

The IRSX ASIC is an 8-channel multi-gigasample per second waveform sampler. It is fabricated in a 0.25 μm complementary metal-oxide-semiconductor (CMOS) process by the Taiwan Semiconductor Manufacturing Company (TSMC).

Each IRSX input channel uses one switched capacitor array (SCA). SCA-based devices [13] have been used in a number of high energy physics readout systems, since they allow large-scale, low-power and wide dynamic range transient signal acquisition. The basic unit of the SCA is a sample and hold analog storage cell, a circuit that has a 14 fF capacitor and a comparator. When an analog switch for the input signal is closed, the input signal is stored in the capacitor (Fig. 9). The analog sample information is kept in the capacitor until the charge is overwritten or until a discharge occurs over time through a leakage current. The charge stored in the capacitor can then be selectively digitized at a later point in time.

Each IRSX channel has a memory depth of 32,768 analog storage cells, internally organized in 16 rows and 32 columns, each addressing a continuous segment of 64 samples.

During TOP operation, a delay locked loop (DLL) circuit fixes the IRSX sampling speed to a submultiple of the SuperKEKB accelerator clock, which is distributed to all Belle II sub-detectors. Analog sampling is carried out continuously with an operational speed of 2.714 GSa/s, resulting in a signal buffer depth of around 12 μs. In addition to the sampling circuits, each channel has a comparator circuit that fires when its input signal exceeds a positive threshold voltage, which is set by an on-die digital-to-analog converter (DAC). Only groups of samples in which these channel trigger outputs coincide with the reception of an external trigger signal from the Belle II Global Decision Logic (GDL) trigger system are digitized and read out. The latency of the Belle II GDL system of 5 μs is significantly shorter than the buffer depth.

After a trigger arrives from the Belle II trigger system, Wilkinson analog-to-digital conversion (ADC) of the waveform samples (Fig. 10) is performed in parallel on the 64 samples corresponding to a single row/column memory read address, simultaneously on all 8 channels. This high degree of parallel processing compensates for the relatively long conversion time required when using the Wilkinson technique; for Belle II operation the conversion time is set to about 4 μs. When a trigger is received, overwriting the respective sample storage segment is blocked by the firmware memory logic so that the stored voltage values are not overwritten until digitization is completed. A common voltage ramp is connected to the positive input of the comparator in every storage cell selected for conversion. For TOP operation, the minimum ramp voltage is set to 0.5 V, while the maximum ramp voltage is set to 2.0 V. At the start of a conversion, the voltage ramp increases linearly from the minimum voltage up to the maximum voltage, during which time an 11-bit Gray code counter is incremented. For a given storage cell, when the voltage ramp level exceeds that of the stored sample voltage, the comparator changes state and latches the Gray code value. In addition, the phase of the Gray code counter clock is latched to provide a 12th bit of ADC resolution. By this method, the voltage ramp and comparator convert the stored voltage into a time interval, and the latched Gray code converts this time into an ADC output. With a 12-bit ADC range, corresponding to the 1.5 V range of the voltage ramp, one ADC least-significant bit corresponds to about 0.4 mV. The Wilkinson ramp, the comparators, and the digital counters are internal to the IRSX ASIC.

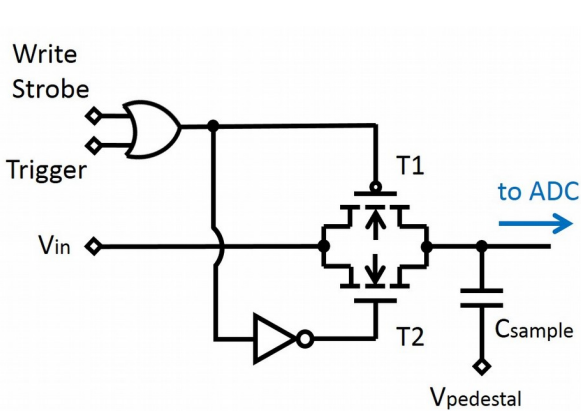

*Figure 9: First stage of a sample and hold storage cell.*

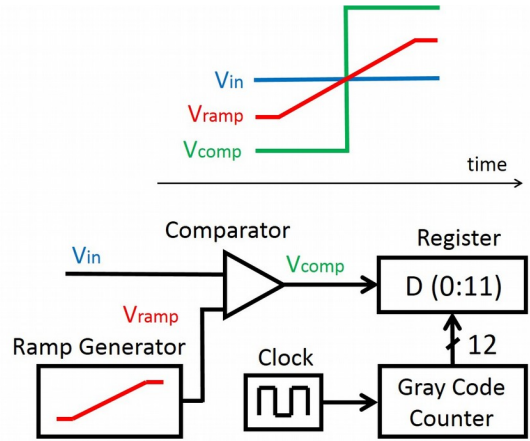

*Figure 10: Wilkinson analog-to-digital conversion process performed on a single sample.*

5. ASIC Carrier Board

Four IRSX chips (a pair on each surface) are mounted on every ASIC Carrier Board. Two spring loaded pogo pin assemblies[1] are installed at one of the edges of the Carrier Board. The pogo pins (one for each MCP-PMT anode) conduct the anode signals for amplification and further sampling and digitization by the IRSX ASICs.

The MCP-PMT anode signal is terminated with a 69.8-ohm terminating resistor, capacitively coupled, and then amplified through a two-stage amplification chain before entering one of the IRSX channels. A pedestal voltage of about 1.0 V is added in the amplification stages to bring the 0 V referenced MCP-PMT signal within the sampling-and-digitization range of the IRSX (0.5 V to 2.0 V). An external calibration pulse input on the SCROD is fanned out to all ASIC Carrier Boards, where it can be routed to inject signals into individual ASIC channels for calibration and testing.

Figure 11 shows a comparison of typical waveforms sampled directly from an MCP-PMT with a high-speed oscilloscope, sampled after the two-stage amplification on the ASIC Carrier board as well as a typical waveform recorded with the IRSX chip on the TOP front-end.

---

1: A custom lower force version of the pogo pins 0926-1-15-20-72-14-11-0 from Mill-Max Manufacturing Corporation.

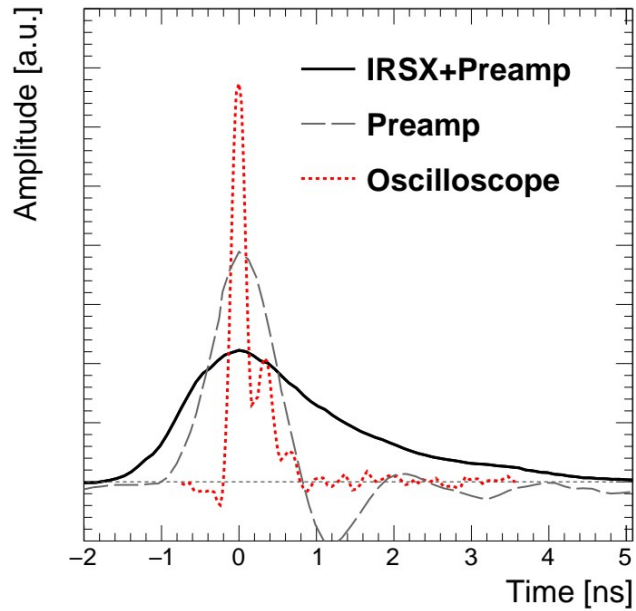

*Figure 11: Typical MCP-PMT pulses as recorded with a fast oscilloscope directly from the sensor (dotted red), after the preamplifier chain installed on each ASIC Carrier (dashed gray) and as sampled by the IRSX ASIC in the TOP system (solid black). The shown pulses are independent measurements that were individually aligned in time and scaled in amplitude for this visualization.*

The amplification chain was optimized for efficient triggering and sampling of MCP-PMT single photon signals with a photomultiplier gain of no less than $2 \times 10^5$. More than 95% of all channels on all installed TOP modules have greater than 80% detection efficiency for single photon PMT pulses. This quoted trigger efficiency is rather conservative, as around half of the MCP-PMTs were operated below the specified PMT gain of $2 \times 10^5$ during these measurements. The nominal PMT gain during operation of the TOP system is $5 \times 10^5$.

Each ASIC Carrier Board houses one Xilinx Zynq Z-7030 system on chip (SoC) that is connected to all four IRSX chips and manages their configuration, operational steering and the readout of digitized waveform segments from the IRSX ASICs. The ASIC Carrier Board has mezzanine connectors that allow the interconnection of up to four Carrier Boards. The connectors support Serializer/Deserializer (SerDes) communication, transmission of low-voltage differential signaling (LVDS) control signals, and transmission of digital data with gigabit per second speeds.

6. SCROD Board

In each SRM, a stack of four interconnected Carrier Boards is electrically and mechanically coupled to a single SCROD Board. The primary element of the SCROD Board is a Xilinx Zynq Z-7045 SoC that receives and aggregates the waveform data recorded by the Carrier SoCs. Upon the receipt of the data, the SCROD SoC processes the waveform data and extracts the time at which the photon hit a particular MCP-PMT pixel. The SCROD also calculates an estimate of the charge collected from the pixel of interest. The extracted hit times and charge estimates are called "feature extracted" data. The controller assigns the feature extracted data an event number and forwards the data to the Belle II DAQ system [14]. For bandwidth reasons, only the feature extracted data, but no waveforms, are transmitted out of the TOP system during standard operations. At the full Belle II trigger rate of 30 kHz, the TOP output bandwidth to the Belle II DAQ will be up to 120 MB/s, for a maximum average data size of around 4 kB per TOP event.

Clock and triggering signals arrive at the SCROD from the Belle II timing distribution system [15] and are sent to each of the four Carrier Board FPGAs through a signal fan-out. The digital data, recorded by the Carrier FPGAs, are sent back to the SCROD through a gigabit per second link realized with mezzanine connectors. The SCROD Board is equipped with two Avago AFBR-57D7APZ small form-factor pluggable fiber optical transceiver modules. One module sends the TOP readout data to the Belle II DAQ system, the other transmits trigger bits from the IRSX ASICs to the Belle II trigger system. Both transceivers are operated at 2.544 gigabits per second, a multiple of the distributed accelerator clock.

The SCROD also has a double data rate random access memory (RAM) to store ASIC parameters and to buffer incoming data. Firmware downloads for the controller and Carrier SoCs can be done through either of two Joint Test Action Group (JTAG) mounted connectors. Both JTAG connectors are accessible for debugging purposes in laboratory settings, but only one connector is used during TOP operation when the SRM is integrated in the detector.

7. High voltage board

In addition to the SCROD and Carrier Boards, each SRM contains one high voltage (HV) board, delivering separate high voltages to each of its eight connected MCP-PMTs. The MCP-PMT operating voltages range from about 2100 V to about 3200 V. Each HV channel is designed as a 400-megaohm resistive divider, coupled with high voltage transistors. The board delivers high voltages to the photocathode and to the microchannel plates of the MCP-PMT. The board is enclosed in an aluminium case, to which the SRM is mechanically attached. Each HV board enclosure is thermally coupled to a water cooled aluminium plate that provides cooling to the whole SRM.

8. Heat removal from the readout modules

The power dissipated in one SRM can be as large as 80 W. Among its components, the main power consumers are the Zynq SoCs, the IRSX ASICs, the signal operational amplifiers in the signal chain, and the voltage regulators. To properly remove heat from the SRM, all ASIC Carrier and SCROD Boards are attached to aluminum plates. The plates are machined to make thermal contact with the highest power dissipation components on the SCROD and ASIC Carrier while leaving the other electrical components unobstructed. Copper disks are epoxied to the Carrier SoCs. A thermal gap filler paste is placed between the copper disks and the aluminum plates to improve the disk-to-plate coupling and improve the heat removal efficiency. During the assembly of an SRM, the edges of the ASIC Carrier and SCROD Boards are sandwiched between the edges of the aluminum plates. In order to increase the heat flow from one plate to another, the edges of each board have a dense array of stitching vias that are filled with thermal grease. Heat is extracted out of the SRMs by the water cooled aluminium base attached to the high voltage board.

9. Front Board

Custom Front Boards (see Fig. 5) couple the MCP-PMTs to the SRM. A 4 x 2 array of MCP-PMTs, corresponding to one SRM and to 128 TOP channels, are served by two Front Boards. The sensors plug into pin sockets installed in blind holes on one side of the Front Board, which are connected to soldering pads on its back side.
When the SRM is installed in a TOP module, the ASIC Carrier pogo pins are pressed against the pads that are connected to the MCP-PMT anode sockets of the Front Board, connecting the MCP-PMT pins with the SRM assembly. Similarly, the outputs of the high voltage board are transferred to the corresponding MCP-PMT pins via pogo pins on the Front Board.

10. Firmware

The firmware of the TOP readout system is implemented on the Xilinx Zynq SoCs assembled on the Carrier and SCROD Boards. Each SoC combines a Kintex-7 field-programmable gate array (FPGA), and a dual-core ARM Cortex-A9 processor with integrated memory into a common assembly package. The TOP firmware is divided into two largely independent parts. One part coordinates the readout, processing and transmission of waveform data from the readout ASICs to the Belle II DAQ system. The other part transmits trigger information to the off-detector TOP trigger system that eventually feeds into the Belle II trigger system.

The FPGA firmware of the Carrier SoC is continually monitoring the channel trigger outputs of all four connected IRSX ASICs. Regions of interest in the IRSX sample buffers are identified by correlating the timing of the received global trigger signals and the recorded channel trigger information. The Carrier FPGA then instructs the readout ASICs to digitize a set number of samples around each possible photon hit, reads out these waveform segments and transfers the raw waveform data to the SCROD Board via SRM-internal serial links. Independent of any global triggers, the Carrier FPGA also continuously streams out the channel trigger information to the SCROD Board. The ARM processor in the Carrier SoC is currently not involved in the readout data flow and only performs low duty cycle slow control tasks.

The SCROD SoC FPGA gathers the waveform data for a given event from all connected Carriers and internally transfers them to its integrated ARM processor. The SCROD ARM processor performs pedestal subtraction for each received waveform sample and extracts the timing of photon signals contained in the waveforms with a constant fraction discrimination algorithm. A data packet of all extracted photon timings, and potentially waveforms for debugging purposes, is then internally transferred back into the SCROD FPGA. The SCROD FPGA adds a packet header containing the module address and trigger number and transfers the data packet to the off-detector Belle II DAQ system via an optical serial link of around 20m length. The SCROD FPGA also receives the trigger streams from all connected Carriers, buffers and sorts them and streams them out to the TOP trigger system on a separate optical transceiver.

At higher background rates, it is foreseen that the constant fraction discrimination algorithm will be moved from software running on the ARM processors into FPGA logic, possibly on the Carrier, to allow for fast parallel processing of waveforms. It is also foreseen to improve the low amplitude photon timing extraction by implementing a template fit algorithm using dedicated digital signal processing resources in the SCROD FPGA.

11. Interface with the Belle II data acquisition system and with the Belle II timing system

A specialized Common Pipelined Platform for Electronic Readout (COPPER) version III (Fig. 12) was developed for the Belle II experiment as a standard component of the DAQ system [13]. The COPPER-III is a 9U Versa Module Europa (VME) platform, based on a 32-bit 1.6 GHz Intel Atom Z530 processor. Four High Speed Link Boards (HSLBs) installed on the COPPER-III board receive data through gigabit per second optical links from four TOP SRMs. The data collected from the TOP module are forwarded to the Belle II event building system [14].

A copy of the 508.9 MHz SuperKEKB radio-frequency clock, divided by four, is provided to all Belle II detector electronics including the TOP SRMs through the Belle II front-end timing switch (FTSW) system [15]. For this purpose, dedicated 6U VME form factor FTSW Boards based on Virtex-5 LX FPGAs were designed and fabricated. The jitter of this Belle II global clock/trigger distribution system is measured to be in the range of 20-30 ps.

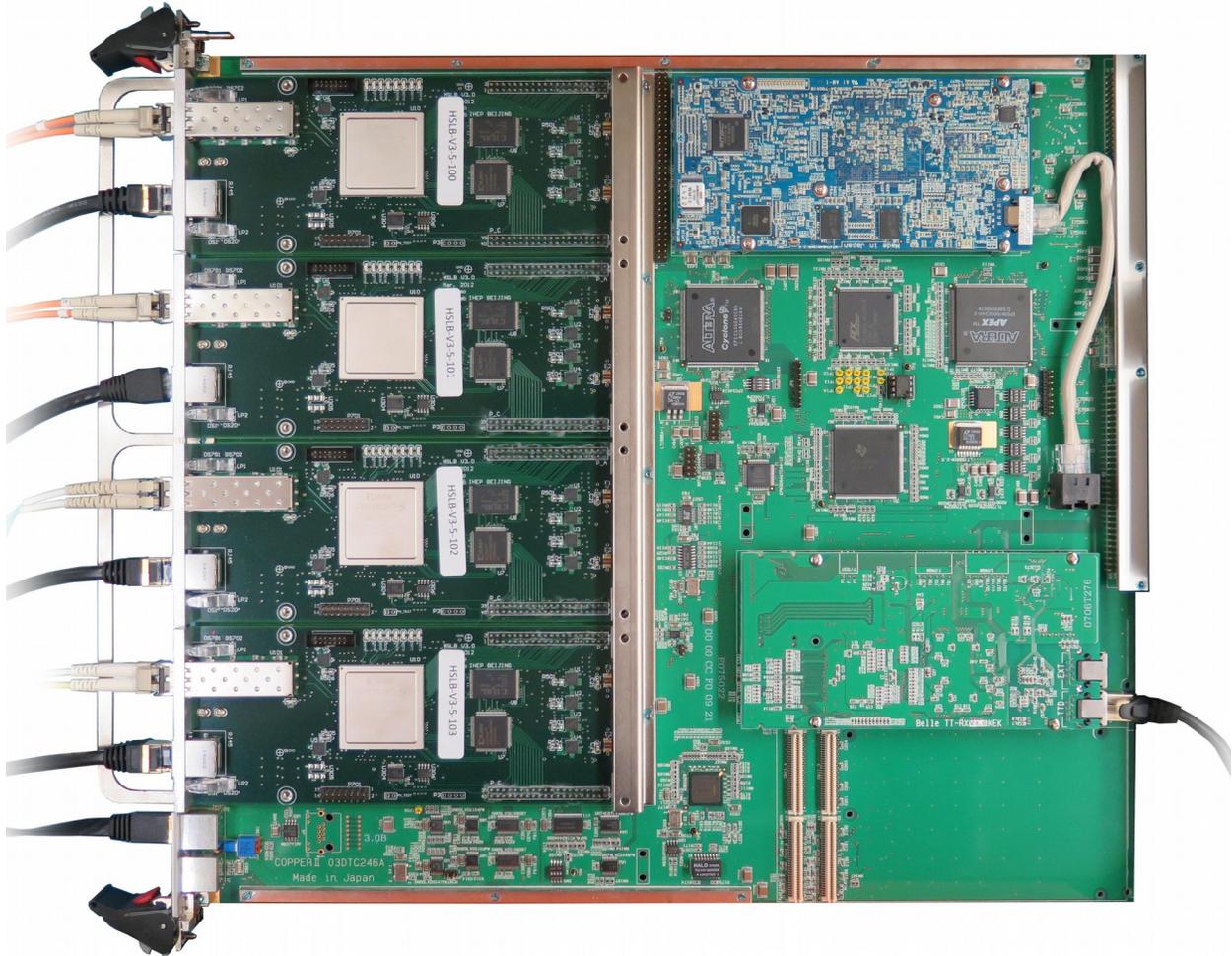

*Figure 12: Common Pipelined Platform for Electronic Readout version III (COPPER-III) with four installed HSLB cards (left), one TTRX mezzanine card (bottom right) and one Intel Atom CPU card (top right, blue PCB). The COPPER PCB has the standard dimensions of a 9U VME card 400mm x 360mm.*

## 12. Performance evaluation of the TOP readout system

The performance of all produced ASIC Carrier Boards and SCROD Boards was evaluated by a variety of measurements in three qualification campaigns. During the first performance evaluation campaign, all ASIC Carrier Boards were tested individually. For each test, the front-end readout system was composed of a single Carrier Board attached to a SCROD Board. The same SCROD Board was used through all measurements, while all of the ASIC Carrier Boards, to be used in the SRMs, were evaluated.

For each channel of each ASIC Carrier Board, various qualification measurements were obtained. Amplitudes and root-mean-square deviations of the pedestal signals were checked to be within specifications. All channel trigger thresholds were characterized and verified. Register reads and writes of the key IRSX parameters that control the sampling, as well as digitization of a fixed amplitude 160 MHz sine wave signal were performed to verify the correct operation of the sampling readout. To verify the expected time resolution of each ASIC, double pulses with a rise time below the bandwidth limit of the on-board preamplifier chain and a fixed temporal separation of 20 ns were injected into each channel. The results demonstrate that the time resolution of the IRSX readout is in the range of 20-30 ps, well within specifications (Fig. 13).

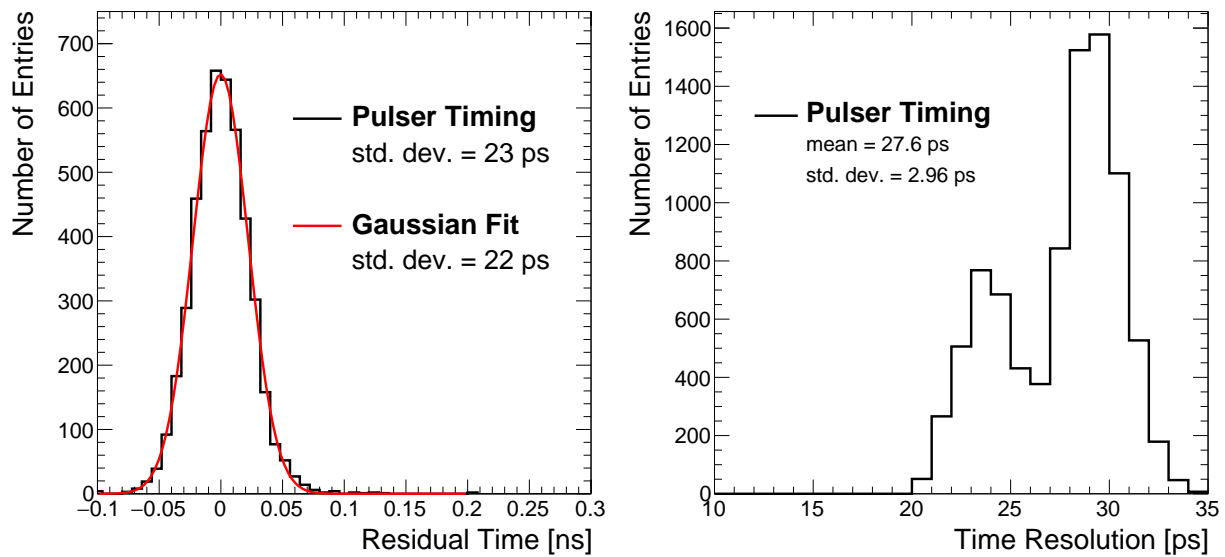

*Figure 13: Single channel electronics timing performance from a measurement of two calibration pulses delayed by 20 ns. Distribution of time deviation measurements for a typical channel (left). Distribution of electronics time resolutions for all channels in all installed TOP SRMs (right). The two separate peaks correspond to characterisation measurements taken in separate test stands.*

The gains of the two-stage amplifier chains for each channel of the ASIC Carrier Board were also measured to ensure that they were within predefined specifications. Those tests were done without MCP-PMTs, using a custom pulse emulator board, which had contact pads of a similar geometry to those of the Front Boards. During the measurements, the Carrier Board pogo pins were pressed against the pads on the emulator board in the same way as they are pressed against the pads on the Front Board when the Carrier is a part of the SRM installed in the TOP (Fig. 14). The gains were measured when the emulator board pad contacts had voltages similar to those of the MCP-PMT anode signal voltages on the pad contacts of the Front Board.

The second performance evaluation campaign included testing firmware programming of all SCROD Boards, together with access tests of their on-board memory. In addition, the quality of all optical transceivers mounted on the SCROD Boards was checked.

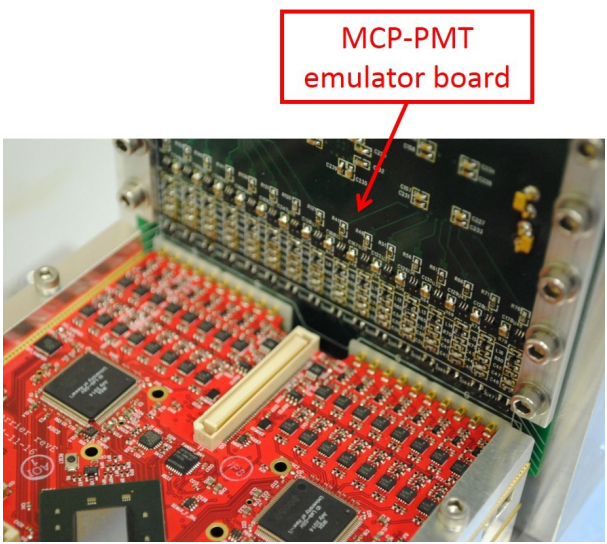

*Figure 15: Reading out emulated MCP-PMT signals with a single ASIC Carrier Board.*

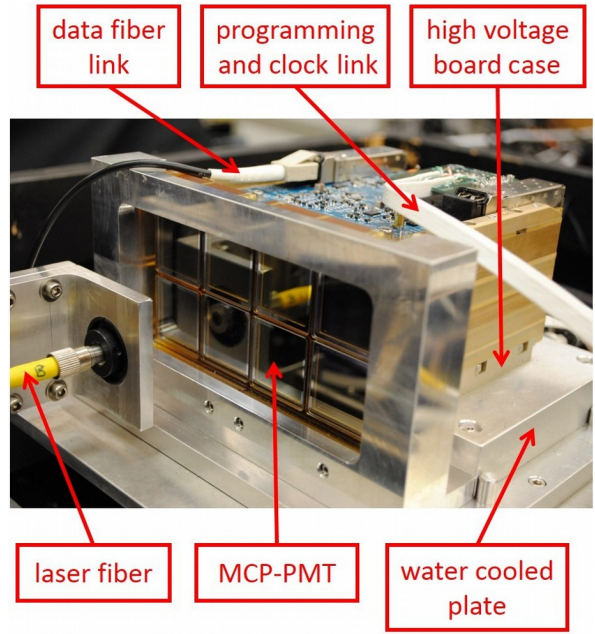

*Figure 14: SRM on the laser test bench. The High Voltage Board is not visible.*

In the third performance evaluation campaign, the fully assembled SRMs were tested at the laser test bench (Fig. 15). The test simulated data taking for the SRM as it would be installed in the TOP system. The laser was tuned for single photon detection by a single MCP-PMT pixel with the MCP-PMT gain in the range from $2 \times 10^5$ to $3 \times 10^5$. A 127.216 MHz global clock was provided to the SRM by one FTSW Board. The laser was asynchronously triggered by an 800 Hz clock. The timing between the leading photon pulse edge and the preceding global clock edge was compared between the sampled IRSX signal and an external 25 ps/count resolution time-to-digital converter (CAMAC Phillips Scientific 7186), yielding a combined time resolution for single photon signals in the TOP front-end electronics. The expected time resolution of a given readout channel coupled to an MCP-PMT anode is the quadratic sum of the IRSX channel timing resolution, the MCP-PMT transit time spread of 30 ps to 40 ps, the FTSW Board clock jitter in the range from 20 ps to 30 ps [15], and the 25 ps/count resolution of the time-to-digital converter used at the test bench. The combined time resolution of the TOP readout including the intrinsic MCP-PMT transient time spread was found to be in the range from 60 ps to 90 ps (Fig. 16).

13. Radiation hardness of the TOP system

The TOP readout electronics must successfully sustain the radiation loads during the operational lifetime of Belle II, which is expected to be at least 10 years. It was estimated that, while operating at the SuperKEKB instantaneous design luminosity of $8 \times 10^{35}$ cm$^{-2}$s$^{-1}$ for 10 years, Belle II will accumulate a total fluence of 1 MeV equivalent neutrons of $15 \times 10^{10}$ n/cm$^2$ and a total radiation dose of 50 Gy in the volume of the TOP detector [16]. To verify their radiation hardness, the SCROD and ASIC Carrier Boards were tested at the Radiation Standards and Calibration Laboratory facilities at the Pacific Northwest National Laboratory [17, 18]. The tested readout system consisted of one Carrier Board attached to the SCROD Board (a configuration similar to that used during the first evaluation campaign described in the previous section). Initially, the test system was exposed to a flux of neutrons from a Cf-252 source, with fluence ranging from $1.5 \times 10^{11}$ n/cm$^2$ to $3.6 \times 10^{11}$ n/cm$^2$. The system was then subjected to a gamma ray dose ranging from 49 Gy to 51 Gy from a Co-60 source. The tested system operated continuously during the irradiation and was monitored for loss of communication, errors in data acquisition program configuration, and changes in voltage and current draw in the SCROD and ASIC Carrier Boards. A constant low rate of recoverable single event upsets was observed during the neutron irradiation. A gradual but permanent increase in the board current draw by less than 5%, with no accompanying faults, was induced by the gamma ray exposure. In both cases, no serious permanent damage was incurred by the on-board components. From these studies it was concluded that the SCROD and ASIC Carrier Boards will function in the radiation environment of Belle II over its lifetime. However, it was also estimated that the TOP readout system will have approximately 70 ± 23 radiation induced single event upset errors in one Belle II operation year equal to $10^7$ seconds at full design luminosity.

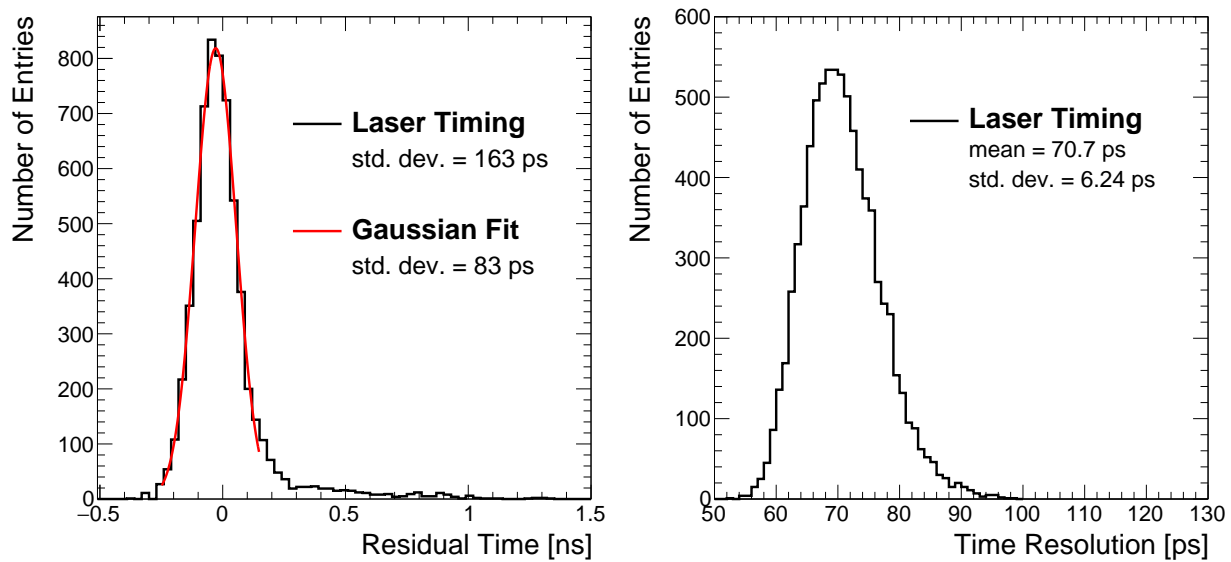

Figure 16: Single photon MCP-PMT timing performance from measurements of fully assembled TOP SRMs on the laser test bench. Distribution of single photon transient time measurements for a typical channel (left). The tail on the right side of the distribution is due to photoelectron backscattering effects in the MCP-PMT. Distribution of single photon time resolutions for all channels on all installed TOP SRMs (right).

14. Summary and future perspectives

In total, 78 front-end Subdetector Readout Modules were assembled from the fabricated SCROD and ASIC Carrier Boards, 64 of which were installed in the TOP detector. The remaining 14 Subdetector Readout Modules are kept as spares. The installation of the full TOP system into the Belle II detector was finished in May 2016. The initial commissioning of the TOP readout system included in situ data taking using a calibration laser as well as cosmic ray muon events with and without the 1.5 T magnetic field of the Belle II solenoid. The data taking demonstrated that the performance of the TOP front-end readout electronics is comparable to or surpasses their performance during the evaluation campaigns. To date, operation of the TOP data acquisition indicates that the SRMs are robust, as the reconstructed data show no signs of degradation in the performance of the readout system.

Results with calibration signals in the installed TOP modules with the full Belle II timing system give electronics time resolutions in the 30-40 ps range for most of the 8192 channels. Further developments of online and offline calibration algorithms to maintain the electronics time resolution in various running conditions are underway.
The TOP detector has been successfully operated in the first collision physics runs of the Belle II experiment from April to July 2018. Detailed studies on calibrations, alignment, time resolution, system stability etc. are ongoing, but analyses of early data with preliminary calibrations have already proven the capabilities of the TOP detector to correctly identify the species of incident charged hadrons.

Acknowledgments

We thank Roy Tom and Curtis McLellan of the University of Hawaii for their work on fabricating the SRM heat removal components and on fabricating testing tooling; Marc Rosen of the University of Hawaii for his work on building the laser test bench; Steven Covin of the University of Hawaii for his work on the SRM assembly; Louis Ridley, Larry Ruckman, and Robin Caplett of the University of Hawaii for their work on designing printed circuit boards for early versions of the TOP readout system; Christina Yee, Casey Honniball, and Grace Jung of the University of Hawaii for their work on soldering the printed circuit boards for the early versions of the TOP readout system; Xin Gao, Sergey Negrashov, Chester Lim, and Andrew Wong for their work on writing and verifying data acquisition firmware and software for the early versions of the TOP readout system; Seth Roffe, Nathan Herring, and Gregory Suehr of the University of Pittsburgh for their work on the performance evaluation of the SCROD Boards; Joe Rabel, George Zuk, and David Emala of the University of Pittsburgh for their work on fabricating low voltage supply cables and custom connectors that interface the SCROD and the FTSW Boards; Richard Lundy of the Pacific Northwest National Laboratory for his work on the pogo pin assembly installation; Craig Bookwalter of the Pacific Northwest National Laboratory for his work on authoring data acquisition code and on performance evaluation of a TOP prototype at a test beam; John Vanderwerp of Indiana University for his work on fabricating the Front Boards; Michael Lang of Indiana University for his work on fabricating the High Voltage Boards; Marko Petric of Jožef Stefan Institute for his work on developing a Geant4 model of Cherenkov photon propagation in the TOP module quartz components. We also thank all members of the Belle II TOP/bPID detector group who worked on installing the SRMs in the TOP modules on-site at KEK, on commissioning the TOP readout system, on taking calibration data, on developing online and offline calibration software, and on monitoring the TOP readout system. This work was supported by the U.S. Department of Energy contracts DE-AC05-76RL01830, DE-SC0012047, DE-SC0010504, DE-SC0010073, and DE-SC0007914. Pacific Northwest National Laboratory is managed and operated by the Battelle Memorial Institute.

References

[1] T. Abe, et al., Belle II Technical Design Report, KEK-Report-2010-1, 2010.
[2] M. T. Cheng, et al., A Study of CP Violation in B Meson Decays: Technical Design Report, KEK Report 95-1, 1995.
[3] A. Abashian, et al., The Belle detector, Nucl. Instr. and Meth. A, 479 (2002) 117-232.
[4] D. M. Asner, et al., US Belle II Project Technical Design Report, available at http://belleweb.pnnl.gov/forTDRreview/TDR-SLAC-13Dec8.pdf (2013).
[5] K. Nishimura, et al., An imaging time-of-propagation system for charged particle identification at a super B factory, Nucl. Instr. and Meth. A, 623 (2010) 297-299.
[6] L. Ruckman, et al., Development of an Imaging Time-of-Propagation (iTOP) prototype detector, Nucl. Instr. and Meth. A, 623 (2010) 365-367.
[7] K. Nishimura, The time-of-propagation counter for Belle II, Nucl. Instr. and Meth. A, 639 (2011) 177-180.
[8] Kenji Inami, TOP counter for particle identification at the Belle II experiment, Nucl. Instr. and Meth. A, 766 (2014) 5-8.
[9] K. Inami, et al., Cross-talk suppressed multi-anode MCP-PMT, Nucl. Instr. and Meth. A, 592 (2008) 247-253.
[10] S. Hirose, et al., Development of the micro-channel plate photomultiplier for the Belle II time-of-propagation counter, Nucl. Instr. and Meth. A, 787 (2015) 293-296.
[11] G. S. Varner, et al., The large analog bandwidth recorder and digitizer with ordered readout (LABRADOR) ASIC, Nucl. Instr. and Meth. A, 583 (2007) 447-460.
[12] G. Varner, et al., The first version buffered large analog bandwidth (BLAB1) ASIC for high luminosity collider and extensive radio neutrino detectors, Nucl. Instr. and Meth. A, 591 (2008) 534-545.
[13] S. A. Kleinfelder, Development of a switched capacitor based multi-channel transient waveform recording integrated circuit, IEEE Trans. Nucl. Sci., 35 (1988) 151-154.
[14] M. Nakao, et al., Data acquisition system for Belle II, JINST, 5 (2010) C12004.
[15] M. Nakao, Timing distribution for the Belle II data acquisition system, JINST, 7 (2012) C01028.
[16] T. Nanut, "TOP beam background", November 2014 Belle II General Meeting, available at https://indico.phys.hawaii.edu/getFile.py/access?contribId=0&resId=0&materialId=slides&confId=1223 (2014).
[17] D. E. Bihl, et al., "Radiation and Health Technology Laboratory Capabilities", PNNL-10354 Rev. 2, available at http://www.pnl.gov/main/publications/external/technical_reports/PNNL-10354rev.2.pdf (2005).
[18] B. G. Fulsom and L. S. Wood, "Radiation hardness testing of iTOP SCROD and carrier boards (v.1.0 – Jan 30 2015)", BELLE2-NOTE-0040, available at https://indico.phys.hawaii.edu/getFile.py/access?contribId=1&resId=0&materialId=0&confId=1223 (2015).